\documentclass[11pt]{article}

\usepackage{graphicx,amsmath,amsthm,mathrsfs,amssymb}
\usepackage[usenames]{color}
\usepackage{ulem}
\usepackage{authblk}
\usepackage{cite}
\usepackage{float}

\def\ltwid{\mathrel{\raise.3ex\hbox{$<$\kern-.75em\lower1ex\hbox{$\sim$}}}}
\def\gtwid{\mathrel{\raise.3ex\hbox{$>$\kern-.75em\lower1ex\hbox{$\sim$}}}}

\def\overleftrightarrow#1{\vbox{\ialign{##\crcr
     $\leftrightarrow$\crcr\noalign{\kern-1pt\nointerlineskip}
     $\hfil\displaystyle{#1}\hfil$\crcr}}}

\newcommand{\be}{\begin{equation}}
\newcommand{\ee}{\end{equation}}
\newcommand{\bea}{\begin{eqnarray}}
\newcommand{\eea}{\end{eqnarray}}
\newcommand{\nn}{\nonumber}

\newcommand\osq{\overline{\square}}
\newcommand\ovr{\overline{R}}

\newcommand\ovGplus{\overline{G}^{+}}
\newcommand\Gplus{G^{+}}

\newcommand{\GN}{G_{\rm N}}

\newcommand{\sbar}{\bar{\sigma}}

\newcommand{\hb}{\bar{h}}
\newcommand{\fb}{\bar{f}}
\newcommand{\fpb}{\bar{f}'}

\newcommand{\dx}[1]{\text{d}#1}
\newcommand{\dd}{\text{d}}

\begin{document}

\begin{titlepage}



\vskip 2cm

\begin{center}
{\bf Does nonlocal gravity yield divergent gravitational energy-momentum fluxes?}
\end{center}

\vskip 1cm

\begin{center}
Yi-Zen Chu$^{1,2\star}$ and Sohyun Park$^{3\dagger}$
\end{center}

\vskip .5cm

\begin{center}
\it{$^{1}$ Department of Physics, National Central University,\\
Chungli 32001, TAIWAN}
\end{center}

\begin{center}
\it{$^{2}$ Center for High Energy and High Field Physics (CHiP), \\
National Central University, Chungli 32001, TAIWAN}
\end{center}

\begin{center}
\it{$^{3}$ CEICO, Institute of Physics of the Czech Academy of Sciences, \\ 
Na Slovance 2, 18221 Prague 8 CZECH REPUBLIC}
\end{center}

\vspace{1cm}

\begin{abstract}
Energy-momentum conservation requires the associated gravitational fluxes on an asymptotically flat spacetime to scale as $1/r^2$, as $r \to \infty$, where $r$ is the distance between the observer and the source of the gravitational waves. We expand the equations-of-motion for the Deser-Woodard nonlocal gravity model up to quadratic order in metric perturbations, to compute its gravitational energy-momentum flux due to an isolated system. The contributions from the nonlocal sector contains $1/r$ terms proportional to the acceleration of the Newtonian energy of the system, indicating such nonlocal gravity models may not yield well-defined energy fluxes at infinity. In the case of the Deser-Woodard model, this divergent flux can be avoided by requiring the first and second derivatives of the nonlocal distortion function $f[X]$ at $X=0$ to be zero, i.e., $f'[0] = 0 = f''[0]$. It would be interesting to investigate whether other classes of nonlocal models not involving such an arbitrary function can avoid divergent fluxes. 
\end{abstract}

\begin{flushleft}
PACS numbers: 04.50.Kd, 95.35.+d, 98.62.-g
\end{flushleft}

\vskip .5cm

\begin{flushleft}
$^{\star}$ e-mail: yizen.chu@gmail.com\\
$^{\dagger}$ e-mail: park@fzu.cz
\end{flushleft}

\end{titlepage}

\section{Introduction}

After more than one hundred years since its formulation, Einstein's General Relativity (GR) is still regarded as the best theory of gravity. GR is based on profound theoretical guidelines such as the equivalence principles, and is supported by a number of experimental tests ranging from millimeter scale in the laboratory to the solar system scales \cite{Will:2014kxa}.
On larger scales, however, GR requires a hypothetical ``dark'' substance, the so-called dark energy to explain the observed acceleration of the universe \cite{SNIA-Riess, SNIA-Perlmutter, newSN, data1, data2, data3, data4, data5}. A cosmological constant, the simplest form of dark energy,  can derive cosmic acceleration. However it is not understood why the energy density of the cosmological constant is orders of magnitude smaller compared to the expectation from quantum field theory, and why it has a value such that it has become dominant only so recently in cosmic history. These are, respectively,  the fine-tuning and coincidence problems of the cosmological constant \cite{CC-review-Weinberg, CC-review-Martin, CC-review-Burgess, CC-review-Padilla}.
Dynamical scalar field models of dark energy also suffer huge fine-tuning problems \cite{scalar-DE-fine-tuning}.   
This has motivated the development of modifications of gravity on cosmological scales 
in order to generate cosmic acceleration without postulating dark energy \cite{MG-review-CFPS, MG-review-CL, MG-review-JJKT, MG-review-NOO}.

When it comes to modifying gravity, we note a theorem that states   
``The only local, metric-based, generally coordinate invariant and potentially stable class of models are 
 $f[R]$ models in which the Ricci scalar is replaced by some nonlinear function of the Ricci scalar" 
\cite{W-lecture}. The data sets suggest that the expansion history is very close to that of the $\Lambda$CDM model \cite{DES-1708.01530,DES-1708.01538}.
However, the only choice of $f[R]$ that can reproduce the $\Lambda$CDM expansion history is $f[R] = R + 2\Lambda$ \cite{f(R)=R-2L}. The three remaining options are:
adding fields other than the metric to carry part of the gravitational force;
breaking general coordinate invariance; and 
abandoning locality. In this paper, we shall focus on the third option of nonlocal modifications of gravity.
This has been less studied compared to the first and second options, so it needs more work \cite{MG-review-CFPS, MG-review-CL, MG-review-JJKT, MG-review-NOO}. 

The justification for nonlocality comes from quantum field theory, in which nonlocality inevitably arises as quantum loop corrections of massless particles; see, for example \cite{Donoghue-1402.3252}.   
It has been suggested that a nonlocal quantum effective action might derive from fundamental theory through the gravitational vacuum polarization of infrared gravitons vastly produced during primordial inflation \cite{W-review, W-shenzhen}. 
However, since no such derivation is currently available, one may take a phenomenological approach, that is to guess what form of nonlocal actions would do the job of generating an accelerated expansion without dark energy \cite{DW-2007, Barvinsky-2011,MM-2014,VAAS-2017,ABN-1707,Tian-1808}.

One should also ensure any modification must be in a way that does not violate the successes of GR in the solar system regime. Furthermore, any gravity theory describing the universe, which has lasted $13.8$ billion years, must yield stable solutions.

To be specific, we consider a metric-based, coordinate invariant, nonlocal model proposed by Deser and Woodard (DW) \cite{DW-2007}, in which the Hilbert Lagrangian is multiplied by an algebraic function of the inverse scalar d'Alembertian acting on the Ricci scalar,
\be
S_{\rm DW} = -\frac{1}{16\pi G}\int d^4 x \sqrt{-g} \left( R + R 
f\Bigl[\frac{1}{\square} R\Bigr] \right)  \;.
\label{DW-action}
\ee
The invariant action guarantees the conservation of field equations obtained from its variation, whether it is local or nonlocal  \cite{W-review}. The nonlocal scalar  $\Box^{-1} R$ is defined with retarded boundary conditions, which ensures causality \cite{W-review,DW-2007}. 
Moreover,  the nonlocal scalar $\Box^{-1}R$ possesses the desired feature of mimicking the behavior of dark energy.  
It naturally delays the onset of cosmic acceleration to late times because
$\square^{-1} R$ grows very slowly: $R = 0$ in a perfect radiation domination so it stays almost zero until the matter-radiation equality and grows logarithmically in the matter dominated era. Further, the function $f[X]$ can be chosen for negative $X \equiv \square^{-1} R$ to reproduce the $\Lambda$CDM expansion history \cite{DW-2009, EPV-1209, EPVZ-1302},
however no huge tuning is required thanks to the delay of $\Box^{-1}R$. Perturbations around the cosmological background have been worked out \cite{PD-2012,DP-2013,PS-2016} and the result was a suppressed growth, which is in better agreement with the data than the $\Lambda$CDM model \cite{NCA-2017,spark-2017}. 

The nonlocal action (\ref{DW-action}) can be re-cast in a localized form by introducing two auxiliary scalar fields \cite{NO-2007, CMN-2008, Koivisto-0803, Koivisto-0807, Koshelev-2008, NOSZ-1010, Bamba-1104, ZS-1108, FeliceSasaki-1412,ZKSZ-1601}. 
One of the two scalar degrees of freedom turns out to be a ghost field, hence the localized version suffers from a kinetic energy instability. However, the original nonlocal model (\ref{DW-action}) is a constrained version of its localized cousin in which the auxiliary scalars and their first derivatives vanish on the initial value surface \cite{DW-2013,PW-1809.06841}, so it can avoid the kinetic instability. In fact, it has been explicitly checked that the evolution of permitted perturbations does not 
lead to explosive excitation of the ghost mode in the original nonlocal model (\ref{DW-action}) but it does so in the localized version \cite{PW-1809.06841}.

The solar system tests on the DW model (\ref{DW-action}) have been first studied in \cite{Koivisto-0807}, where 
Koivisto has shown that the constraint from the Cassini spacecraft \cite{Cassini} only fixes the first derivative of the nonlocal function $f[X]$ at $X= 0$ to lie within the range  
\be
-5.8 \times 10^{-6} < f'[0] < 5.7 \times 10^{-6}
\label{fbprime-bound}
\ee
and fixes none of the higher derivatives \cite{Koivisto-0807}. Note that, in his analysis,
the nonlocal function was expanded around the Minkowski background, in which $X = \Box^{-1}R = 0$. This bound is easily satisfied by the nonlocal function found to reproduce the $\Lambda$CDM expansion history, $f_{\Lambda\text{CDM}}'[0] \sim 10^{-24}$ \cite{DW-2009}. Deser and Woodard has pointed out that a perfect screening can be made inside the solar system: 
The key point is that the scalar d'Alembertian $\Box$ has different signs when acting on functions of time than on functions of space. 
As a result, $X = \Box^{-1} R < 0$ for cosmological scales and $X = \Box^{-1} R > 0$ for gravitationally bound systems. 
Since the reproduction of the $\Lambda$CDM expansion history fixes the function $f[X]$ for $X < 0$ and $f[0] = 0$  \cite{DW-2009}, defining $f[X] = 0$ for all $X>0$ can completely eliminate any corrections inside gravitationally bound systems \cite{DW-2013}. This choice amounts to setting to zero $f[0]$ as well as all its derivatives $f^{(n \geq 1)}[0]$. 

In the present paper, on the other hand, we do not apply this $f[X \geq 0] = 0$ assumption. Instead, we wish to constrain the nonlocal function $f$ in a more direct way by answering the question: ``What is the gravitational energy-momentum generated by isolated astrophysical sources in asymptotically flat spacetimes?" In GR, the quadrupole radiation formula and its implications for the dynamics of compact binary systems have been well tested since the discovery of the Hulse-Taylor binary pulsar \cite{Hulse:1974eb}. (See \cite{Weisberg:2004hi} for a recent review.) At this point, any modifications of gravity must therefore pass this consistency test at the sub-percent level. What we uncover instead is that, once $f'[0] \neq 0$ and $f''[0] \neq 0$, the total energy-momentum of gravitational waves within the Deser-Woodard model likely diverges at infinity.

The rest of this paper is organized as follows. In Section \ref{Section_Setup}, we lay out the setup, including the field equations and definitions we will use throughout the paper. Section \ref{Section_PerturbativeAnalysis} is the heart of our paper. We first expand the field equations around the flat spacetime background to linear order in metric perturbations and study gravitational polarizations and the non-relativistic/static limit.
We then further expand up to quadratic order in metric perturbations and compute the gravitational energy-momentum flux due to an isolated system. Our discussions comprise Section \ref{Section_Discussion}. 
In Appendices \ref{Section_ExpansionOfNonlocalTerms} and \ref{Section_GreenFunction} we describe, respectively, the expansion of various nonlocal terms occurring within the Deser-Woodard equations-of-motion and the solution of the de Donder gauge retarded Green's function of its linearized metric perturbation. 

\section{Setup: Deser-Woodard (DW) Model}
\label{Section_Setup}

We start from the causal and conserved nonlocal field equations of the DW model derived in \cite{DW-2007}:
\bea
G_{\mu\nu} +\Delta G_{\mu\nu} = 8\pi \GN T_{\mu\nu} ,
\label{DW-EoM}
\eea
where the nonlocal modifications $\Delta G_{\mu\nu}$ to the Einstein tensor $G_{\mu\nu}$ are
\be
\Delta G_{\mu\nu} = \Delta^{A} G_{\mu\nu} +  \Delta^{B} G_{\mu\nu} , 
\ee
with 
\bea
\Delta^{A} G_{\mu\nu} &\equiv&  \left( G_{\mu\nu} + g_{\mu\nu} \Box - \nabla_\mu \nabla_\nu \right) \left( f\left[ \frac{1}{\Box} R \right] +  \frac{1}{\Box} \left( R f'\left[\frac{1}{\Box} R\right] \right) \right) , 
\\
\Delta^{B} G_{\mu\nu} &\equiv&
\left( \frac{1}{2} \delta^{\{\alpha}_\mu \delta^{\beta\}}_\nu 
- \frac{1}{2} g_{\mu\nu} g^{\alpha\beta} \right) \partial_\alpha \left(\frac{1}{\Box} R \right)
\partial_\beta \left(\frac{1}{\Box} \left( R f'\left[\frac{1}{\Box} R\right] \right) \right) .
\eea
Here the symmetrization symbol means 
$
A_{\{\alpha}B_{\beta\}}  \equiv  A_{\alpha}B_{\beta} + A_{\beta}B_{\alpha} .
$
On the right hand side of Eq. \eqref{DW-EoM}, we will let $T_{\mu\nu}$ denote the stress tensor of some astrophysical source of gravitational waves, such as the binary systems detected by LIGO/Virgo to date. We shall assume the spacetime metric is given by the following deviation from the Minkowski one:
\begin{align}
\label{PerturbedMinkowski}
g_{\mu\nu} = \eta_{\mu\nu} + h_{\mu\nu} .
\end{align}
We will work with the metric signature $(+, -, -, -)$. In the following, we will primarily use the `trace-reversed' perturbation variable
\begin{align}
\label{TraceReversed}
\hb_{\mu\nu} \equiv h_{\mu\nu} - \frac{1}{2} \eta_{\mu\nu} \hb, \qquad
\hb \equiv \eta^{\sigma\rho} \hb_{\sigma\rho} .
\end{align}
It is not expected that an exact solution can be obtained, so the strategy is to employ perturbation theory. To this end, we lay out in Appendix \ref{Section_ExpansionOfNonlocalTerms}, the results of expanding 
in power series of $\hb_{\mu\nu}$ the nonlocal terms -- those involving $\Box^{-1}$ -- in Eq. \eqref{DW-EoM}. 
We now proceed to massage Eq. \eqref{DW-EoM} into a power series in the perturbation $\hb_{\mu\nu}$ (or, equivalently, in $h_{\mu\nu}$). If $\delta_n (\dots)$ represents the piece of $(\dots)$ containing precisely $n$ powers of the metric perturbation, then Eq. \eqref{DW-EoM} can be rearranged as 
\begin{align}
\label{DW-EoM_InfiniteSeries}
\delta_1 G_{\mu\nu} + \delta_1 \Delta G_{\mu\nu}
&= 8\pi \GN \tau_{\mu\nu} , \\
\tau_{\mu\nu}
&\equiv \sum_{\ell=0}^{\infty} \delta_\ell T_{\mu\nu} - (8\pi\GN)^{-1} \sum_{\ell=2}^{\infty} \left\{ \delta_\ell G_{\mu\nu} + \delta_\ell \Delta G_{\mu\nu} \right\} .
\end{align}
The results from Appendix \ref{Section_ExpansionOfNonlocalTerms} tell us that the linear-in-$h$ terms on the left hand side of Eq. \eqref{DW-EoM_InfiniteSeries} are
\be
\delta_1 G_{\mu\nu} + \delta_1 \Delta G_{\mu\nu}
= (1+\fb) \delta_1 G_{\mu\nu} + 2 \fpb (\eta_{\mu\nu} \partial^2 - \partial_\mu \partial_\nu) \int \ovGplus[x-x'] \delta_1 R[x'] \dd^4 x',
\label{1st-DW-EoM}
\ee
where $\fb \equiv f[0], \fb' \equiv f'[0]$ and the first order Einstein tensor and Ricci scalar are, respectively,
\begin{align}
\delta_1 G_{\mu\nu} 
&= -\frac{1}{2} \left( \partial^2 \bar{h}_{\mu\nu} - \partial_{\{\mu} \partial^\sigma \bar{h}_{\nu\}\sigma} + \eta_{\mu\nu} \partial^{\sigma} \partial^\rho \bar{h}_{\sigma\rho}  \right) 
\label{1st-Gmm} , \\
\delta_1 R &= \partial^{\sigma} \partial^\rho \bar{h}_{\sigma\rho} + \frac{1}{2} \partial^2 \bar{h} ;
\label{1st-RicciScalar} 
\end{align}
and the retarded Green's function of the wave operator $\partial^2 \equiv \eta^{\alpha\beta} \partial_\alpha \partial_\beta$, obeying
\begin{align}
\label{GreenFunctionEoM}
\partial^2 \ovGplus[x] = \delta^{(4)}[x] ,
\end{align}
is given by the following expression involving the Dirac delta function:
\begin{align}
\label{GreenFunction}
\ovGplus[x] = \frac{\delta[t-|\vec{x}|]}{4\pi|\vec{x}|} .
\end{align}
{\bf Conservation} \qquad The key observation we wish to make is that, via a direct calculation using Eqs. \eqref{1st-DW-EoM} and \eqref{1st-Gmm}, one can readily verify the left hand side of Eq. \eqref{DW-EoM_InfiniteSeries} is conserved with respect to the background flat metric:
\begin{align}
\partial^\mu \left(\delta_1 G_{\mu\nu} + \delta_1 \Delta G_{\mu\nu}\right) = 0 .
\end{align}
In fact, the two terms are separately conserved. In the far zone where the observer is located and the astrophysical stress tensor is zero, this allows us to interpret the gravitational terms quadratic and higher in perturbations on the right hand side of Eq. \eqref{DW-EoM_InfiniteSeries} to be associated with gravitational stress-energy-momentum -- as in \S 7.6 of Weinberg \cite{Weinberg:1972kfs}. As $r \to \infty$, we expect $| \hb_{\mu\nu} | \ll 1$ and the quadratic terms should be the dominant ones in this limit. Hence, we shall identify the far zone gravitational stress-energy tensor as the quadratic terms
\begin{align}
\label{GravitationalStressTensor}
t_{\mu\nu} \equiv -\frac{\delta_2 G_{\mu\nu} + \delta_2 \Delta G_{\mu\nu}}{8\pi \GN} .
\end{align}

\section{DW Model: Perturbative Analysis}
\label{Section_PerturbativeAnalysis}

In this section, we will solve for Eq. \eqref{GravitationalStressTensor} engendered by an isolated astrophysical system via perturbation theory. The usual strategy is to first obtain the solutions to the linearized form of Eq. \eqref{DW-EoM_InfiniteSeries}; the second order solutions may then be obtained via iteration, etc. However, to leading order in $\GN$, we only need to insert the linearized solutions into Eq. \eqref{GravitationalStressTensor} to obtain the leading order $\mathcal{O}[\GN]$ contributions to GW stress energy flux since higher order solutions necessarily yield higher powers of $\GN$ in $t_{\mu\nu}$.

To obtain the linearized solutions to Eq. \eqref{DW-EoM_InfiniteSeries}, we choose the de Donder gauge
\begin{align}
\label{deDonder}
\partial^\mu \hb_{\mu\nu} = 0 .
\end{align}
This simplifies the linearized Einstein tensor and Ricci scalar in Eqs. \eqref{1st-Gmm} and \eqref{1st-RicciScalar} to
\begin{align}
\delta_1 G_{\mu\nu}
= -\frac{1}{2} \partial^2 \bar{h}_{\mu\nu}
\label{1st-Gmm-deDonder}
\qquad \text{ and } \qquad
\delta_1 R
= \frac{1}{2} \partial^2 \bar{h} .
\end{align}
Furthermore, Eqs. \eqref{GreenFunctionEoM} and \eqref{1st-Gmm-deDonder} now inform us
\bea
\int \dx^4 x' \ovGplus[x-x'] \delta_1 R[x']
= \frac{1}{2} \bar{h}[x] ,
\label{1st-Green-integral}
\eea
where we have assumed the wave operator may be integrated-by-parts to act on the Green's function. Altogether, Eqs. \eqref{1st-Gmm-deDonder} and \eqref{1st-Green-integral} applied to Eq. \eqref{DW-EoM_InfiniteSeries} hands us
\be
\label{DeserWoodardModel_deDonder}
-\frac{1}{2} \left( 1 + \fb \right) \partial^2 \hb_{\mu\nu}
+ \fpb \left( \eta_{\mu\nu} \partial^2 - \partial_\mu \partial_\nu \right) \hb
= 8\pi \GN \tau_{\mu\nu} .
\ee
\subsection{Linearized Solutions, Gravitational Polarizations, Static Limit}
In Appendix \ref{Section_GreenFunction} we evaluate the Green's function of the wave operator on the left hand side of Eq. \eqref{DeserWoodardModel_deDonder}. We find that
\begin{align}
\hb_{\mu\nu}[x]
&= -8\pi\GN \int \dd^4 x' \left( \frac{2}{1+\fb} \overline{G}^+[x-x'] \tau_{\mu\nu}[x'] + \frac{4 \fpb}{(1+\fb)(1+\fb-6\fpb)} \text{DW}_{\mu\nu}[x-x'] \tau[x'] \right) , \\
\tau &\equiv \eta^{\alpha\beta} \tau_{\alpha\beta} 
\end{align}
where, for the reader's convenience, we collect the results from Eq. \eqref{GreenFunction} and Appendix \ref{Section_GreenFunction}:
\begin{align}
\overline{G}^+[x-x'] &= \frac{\delta[T-R]}{4\pi R}, \\
\text{DW}_{\mu\nu}[x-x'] 
&= \eta_{\mu\nu} \frac{\delta[T-R]}{8\pi R}
+ \frac{(x-x')_\mu (x-x')_\nu}{R} \frac{\partial}{\partial R} \frac{\delta[T-R]}{8\pi R} \\
&= \eta_{\mu\nu} \frac{\delta[T-R]}{8\pi R}
- \frac{(x-x')_\mu (x-x')_\nu}{8\pi R^2} \left( \delta'[T-R] + \frac{\delta[T-R]}{R} \right) , \\
T &\equiv t-t', \qquad R \equiv |\vec{x}-\vec{x}'| .
\end{align}
More explicitly, we have
{\allowdisplaybreaks\begin{align}
\label{hb_00}
\frac{\hb_{00}}{8\pi\GN}
	&= - \frac{2}{1+\fb} \int \dd^4 x' \frac{\delta[T-R]}{4\pi R} \tau_{00}[x'] 
	+ \frac{4 \fpb}{(1+\fb)(1+\fb-6\fpb)} \int \dd^4 x' \frac{\delta[T-R]}{8\pi} \left( \dot{\tau}[x'] - 2 \frac{\tau[x']}{R} \right) , \\
\label{hb_0i}
\frac{\hb_{0i}}{8\pi\GN} 
	&= - \frac{2}{1+\fb} \int \dd^4 x' \frac{\delta[T-R]}{4\pi R} \tau_{0i}[x'] 
	+ \frac{4 \fpb}{(1+\fb)(1+\fb-6\fpb)} \int \dd^4 x' \delta[T-R] \left( \frac{R_i}{8\pi R} \dot{\tau}[x'] \right) , \\
\label{hb_ij}
\frac{\hb_{ij}}{8\pi\GN} 
&= - \frac{2}{1+\fb} \int \dd^4 x' \frac{\delta[T-R]}{4\pi R} \tau_{ij}[x'] \nonumber\\
& \hspace{0.5cm} - \frac{4 \fpb}{(1+\fb)(1+\fb-6\fpb)} \int \dd^4 x' \frac{\delta[T-R]}{8\pi} \left( -\delta_{ij} \frac{\tau[x']}{R} - \frac{R_i R_j}{R^2} \left(\dot{\tau}[x'] + \frac{\tau[x']}{R} \right) \right) .
\end{align}

{\bf GW Polarizations} \qquad We will employ the de Donder gauge solutions in Eqs. \eqref{hb_00}, \eqref{hb_0i}, and \eqref{hb_ij} to understand the GW polarizations of the DW model. First, the gauge-invariant $0i0j$ components of the linearized Riemann tensor in the de Donder gauge of Eq. \eqref{deDonder} are
\begin{align}
\label{LinearizedRiemann_dD}
\delta_1 R_{0i0j} 
&= \frac{1}{2} \left( \partial_l \partial_{\{i} \hb_{j\}l} - \partial_0^2 \left( \hb_{ij} - \frac{1}{2} \delta_{ij} \hb_{ll} \right) - \frac{1}{2} \delta_{ij} \partial_l \partial_m \hb_{lm} - \frac{1}{2} \partial_i \partial_j \hb_{ll} - \frac{1}{2} \partial_i \partial_j \hb_{00} \right) .
\end{align}
On the other hand, if one chooses instead the synchronous gauge, where all perturbations occur within the spatial metric,
\begin{align}
\label{SynchronousGauge}
\dd s^2 = \eta_{\mu\nu} \dd x^\mu \dd x^\nu + h_{ij}^{\text{(s)}} \dd x^i \dd x^j ,
\end{align}
the linearized Riemann components would read instead
\begin{align}
\label{LinearizedRiemann_Synchronous}
\delta_1 R_{0i0j} 
&= -\frac{1}{2} \ddot{h}_{ij}^{(\text{s})} .
\end{align}
Since GWs are detected in the far zone $r \to \infty$ limit, we shall examine the linearized $\hb$ solutions in this limit. Now, all of them in Eqs. \eqref{hb_00}, \eqref{hb_0i} and \eqref{hb_ij} take the form
\begin{align}
\hb_{\mu\nu} \sim \int \dd^4 x' \delta[T-R] Q_{\mu\nu}^{\phantom{\mu\nu} \alpha\beta} \tau_{\alpha\beta} 
\end{align}
where $Q_{\mu\nu}^{\phantom{\mu\nu} \alpha\beta}$ could potentially involve $R^i/R$ or powers of $1/R$. If we agree to put the center of the spatial coordinate system within the astrophysical system, then in the far zone $R = r - \vec{x}' \cdot \widehat{r} + \mathcal{O}(r \cdot (|\vec{x}'|/r)^2)$, where $r \equiv |\vec{x}|$ and $\widehat{r} \equiv \vec{x}/r$. Since $r^{-1},R^{-1} \to 0$ in the far zone, more spatial derivatives acting on the $Q_{\mu\nu}^{\phantom{\mu\nu} \alpha\beta}$ would produce faster decay in the far zone. Therefore, to leading order, each spatial derivative acting on $\hb_{\mu\nu}$ acts on the $\delta$-function, which in turn produces
\begin{align}
\partial_i \delta[T-R] = -\frac{R^i}{R} \partial_t \delta[T-R] .
\end{align}
But in the far zone, the $R^i/R \to \widehat{r}^i$. Hence, the far zone replacement rule can be summarized as
\begin{align}
\partial_i \hb_{\mu\nu} \to -\widehat{r}^i \partial_0 \hb_{\mu\nu}, \qquad\qquad
\widehat{r}^i = \frac{x^i}{r} .
\end{align}
Moreover, $\partial_i \widehat{r}^j$ would also yield more powers of $1/r$; hence to leading order and using the de Donder gauge condition, $\partial_i \partial_j \hb_{00} \to \widehat{r}^i \widehat{r}^j \partial_0^2 \hb_{00} = \widehat{r}^i \widehat{r}^j \partial_l \partial_m \hb_{lm} \to \widehat{r}^i \widehat{r}^j \widehat{r}^l \widehat{r}^m \partial_0^2 \hb_{lm}$. Altogether, we gather the far zone de Donder gauge linearized Riemann is
\begin{align}
\label{LinearizedRiemann_dD_FarZone}
\delta_1 R_{0i0j}
&\to -\frac{1}{2} \ddot{h}_{ij}^{\text{tt}} ,
\end{align}
where the transverse-traceless gravitational perturbation is
\begin{align}
h_{ij}^{\text{tt}}
&= \left( P_{ia} P_{jb} - \frac{1}{2} P_{ij} P_{ab} \right) \hb_{ab} , \\
P_{ab} &\equiv \delta_{ab} - \widehat{r}_a \widehat{r}_b .
\end{align}
As the name suggests, the GW is traceless and transverse to the direction of GW propagation:  
\be
\delta^{ij} h_{ij}^{\text{tt}} = 0, ~~ \widehat{r}^i h_{ij}^{\text{tt}}=0.
\ee
As already alluded to, the linearized Riemann is gauge invariant. Hence, we may set Eqs. \eqref{LinearizedRiemann_Synchronous} and \eqref{LinearizedRiemann_dD_FarZone} equal in the far zone, and -- by integrating them twice with respect to time -- obtain (up to initial conditions) the equality of the synchronous gauge and the transverse-traceless GW:
\begin{align}
h_{ij}^{(\text{s})} = h_{ij}^{\text{tt}} + \dots .
\end{align}
Because the synchronous gauge has purely spatial perturbations, we may interpret $h_{ij}^{(\text{s})}$ to be a direct measure of the fluctuations in proper lengths of infinitesimally separated test masses co-moving in the spacetime where $t$ refers to their proper times. More specifically, we see this is identical to the corresponding result in General Relativity, except $\GN$ is to be replaced by $\GN/(1+\fb)$:
\begin{align}
h_{ij}^{\text{(s)}} 
\approx - \frac{16\pi\GN}{1+\fb} \left( P_{ia} P_{jb} - \frac{1}{2} P_{ij} P_{ab} \right) 
\int_{\mathbb{R}^3} \frac{\dd^3 \vec{x}'}{4\pi r} \tau_{ab}[t-r+\vec{x}'\cdot\widehat{r},\vec{x}'] + \dots,
\end{align}
because the second line of Eq. \eqref{hb_ij} is eliminated by the `tt' projector. 

In summary, linear GWs in the DW model yield the same signals as in GR -- i.e., far-zone `spin-2' waves sourced by the transverse-traceless portion of the spatial energy-momentum-shear-stress tensor -- except the effective Newton's constant $\GN$ is replaced by $\GN/(1+\fb)$. 
It implies that the GWs still travel at the speed of light though their amplitude is rescaled by $1/(1+\fb)$. Therefore the propagation of GWs in the DW model satisfies the exquisite bound on the speed of GWs
\be
\left|\frac{v_{\text{\tiny GW}}}{c} - 1 \right|\leq 5 \times 10^{-16}
\ee
placed by the detection of the binary neutron star merger GW170817  \cite{GW170817, GRB 170817A}. 
This is in contrast to the modified gravity models of the first option, adding fields other than the metric to carry part of the gravitational force, which typically predict a variable GW speed \cite{Lombriser:2015sxa,Lombriser:2016yzn,McManus:2016kxu,GW-1,GW-2,GW-3,GW-4}.

{\bf Non-relativistic/Static limit} \qquad Next, we move on to examine gravitational tidal forces exerted by isolated non-relativistic systems.

Within this non-relativistic/static limit, we assume $\tau_{ij}$ and $\tau_{0i}$ are negligible compared to $\tau_{00}$ and that $\tau_{00}$ itself is time independent. This is a simplified model for, say, describing the tidal forces the Moon exerts on the Earth. In this limit, we have to leading order in $1/r$,
\begin{align}
\hb_{00}
&\approx - \left( \frac{2}{1+\fb} + \frac{4 \fpb}{(1+\fb)(1+\fb-6\fpb)} \right) \frac{2 \GN M}{r} , \\
\hb_{ij}
&\approx \frac{4 \fpb}{(1+\fb)(1+\fb-6\fpb)} \left( \delta_{ij} + \widehat{r}^i \widehat{r}^j \right) \frac{1}{2} \frac{2 \GN M}{r} , \\
\hb_{0i} &\approx 0, \qquad\qquad
M \equiv \int \dd^3 \vec{x}' \tau_{00}[\vec{x}'] .
\end{align}
In this static limit,
\begin{align}
\delta_1 R_{0i0j} = -\left( \frac{1}{1+\fb} - \frac{2 \fpb}{(1+\fb)(1+\fb-6\fpb)} \right) \left( \delta_{ij} - 3 \widehat{r}_i \widehat{r}_j \right) \frac{\GN M}{r^3} .
\end{align}
Notice the `effective' Newton's constant 
\begin{align}
G_{\text{tidal}} \equiv \left( \frac{1}{1+\fb} - \frac{2 \fpb}{(1+\fb)(1+\fb-6\fpb)} \right) \GN = \frac{1}{1+\fb} \frac{1+\fb-8\fpb}{1+\fb-6\fpb} \GN ,
\end{align}
which determines the strength of tidal forces, is different from that determining the strength of gravitational waves
\begin{align}
G_{\text{GW}} \equiv \frac{\GN}{1+\fb} 
\end{align}
unless $\fb' = 0$.
Moreover, a non-relativistic particle of mass $m \ll M$ near this monopole would have an action -- assuming there are no external forces --
\begin{align}
-m \int \sqrt{1-\vec{v}^2  + h_{00} + \mathcal{O}[v^2,h]} \dd t = - m \int \left( 1 - \frac{1}{2} (\vec{v}^2 - h_{00}) + \dots  \right) \dd t .
\end{align}
This allows us to identify the `Newtonian potential' in $\vec{a}^i = -\partial_i \Psi_\text{N}$ as
\begin{align}
\Psi_{\text{N}} = \frac{1}{2} h_{00} = -\frac{1}{1+\fb} \frac{1+\fb-8\fpb}{1+\fb-6\fpb} \frac{\GN M}{r} .
\end{align}
This is also consistent with the results above for the linearized Riemann tensor, as well as \cite{Koivisto-0807}, since one expects $\delta_1 R_{0i0j} = - \partial_{i}\partial_{j} \Psi_{\text{N}}$.

At this point, we have the following weak field metric
\begin{align}
\dd s^2 
&= \left( 1 - \alpha \frac{r_s}{r} \right) \dd t^2 - \left( 1 + (\beta-\gamma) \frac{r_s}{r} \right) \dd r^2 
- \left( 1 + \beta \frac{r_s}{r} \right) r^2 \Omega^2 \\
r_s &\equiv 2\GN M .
\end{align}
where
\begin{align}
\alpha 	= \frac{1}{1+\fb} \frac{1+\fb-8\fpb}{1+\fb-6\fpb}, \qquad
\beta 	= \frac{1}{1+\fb} \frac{1+\fb-2\fpb}{1+\fb-6\fpb}, \qquad
\gamma	= \frac{1}{1+\fb} \frac{2\fpb}{1+\fb-6\fpb} .
\end{align}
Let us define
\begin{align}
r \equiv \left( 1 - \frac{\beta}{2} \frac{r_s}{r'} \right) r' 
\end{align}
to first order in $r_s/r'$. We discover
\begin{align}
\dd s^2 
&= \left( 1 - \alpha \frac{r_s}{r'} \right) \dd t^2 - \left( 1 + (\beta-\gamma) \frac{r_s}{r'} \right) \dd r'^2 - r'^2 \Omega^2 ,
\end{align}
and also observe that
\begin{align}
\beta - \gamma &= \frac{1}{1+\fb} \frac{1+\fb-4\fpb}{1+\fb-6\fpb} \neq \alpha .
\end{align}
To order $r_s/r$, the Ricci scalar is zero for any $\alpha,\beta,\gamma$; but the Ricci tensor is zero if and only if 
$\alpha = \beta-\gamma$. Therefore, our solution here does not satisfy the vacuum Einstein equations (i.e., GR). Hence if this solution can be regarded as the far zone region of a black hole solution, it would not be the Schwarzschild solution. In other words, the DW model violates the Birkhoff theorem \cite{Birkhoff,Jebsen,Eisland,Alexandrow}. It also implies that the two Newtonian potentials identified as 
\begin{align}
\Psi_{\text{N}}  &= - \alpha \frac{r_s}{2r} = - \frac{1}{1+\fb} \frac{1+\fb-8\fpb}{1+\fb-6\fpb}\frac{\GN M}{r} \\
\Phi_{\text{N}} &= - (\beta-\gamma) \frac{r_s}{2r} = -\frac{1}{1+\fb} \frac{1+\fb-4\fpb}{1+\fb-6\fpb} \frac{\GN M}{r} 
\end{align}
are not equal to each other.
This peculiar behavior can be avoided if one sets $\fb' = 0$.
In fact, Koivisto has given the bound $-5.8 \times 10^{-6} < \fb' < 5.7 \times 10^{-6}$ in Eq. \eqref{fbprime-bound} upon the observation of $\Psi_{\text{N}} \neq \Phi_{\text{N}}$ \cite{Koivisto-0807}.
Note that all the anomalies at linear order such as $G_{\text{GW}} \neq G_{\text{tidal}}$ and $\Psi_{\text{N}} \neq \Phi_{\text{N}}$ would disappear if $\fb' = 0$. 

\subsection{Quadratic Order and $t_{\mu\nu}$[GW]}

At quadratic order, let us focus on the nonlocal terms in the far zone GW stress tensor in Eq. \eqref{GravitationalStressTensor},
\begin{align}
\delta_2 \Delta G_{\mu\nu}
&\equiv \delta_2 \Delta^{A} G_{\mu\nu} + \delta_2 \Delta^{B} G_{\mu\nu} ,
\end{align}
where the modified Einstein tensor terms at second order are 
{\allowdisplaybreaks
\bea
\delta_2 \Delta^{A} G_{\mu\nu}  
&\!\!\!\!\!=\!\!\!\!\!& \delta_2 G_{\mu\nu} \fb+ 2\delta_1 (G_{\mu\nu} +g_{\mu\nu} \Box - \nabla_\mu \nabla_\nu ) \fb' \int \ovGplus \delta_1 R 
+ ( \eta_{\mu\nu} \partial^2 - \partial_\mu \partial_\nu) \Biggl\{
\frac{1}{2}\fb'' \left(\int  \ovGplus \delta_1 R \right)^2
\nn \\ 
&& + \fb'' \int \ovGplus \delta_1 R \int  \ovGplus \delta_1 R  + \fb' \int \ovGplus h \delta_1 R +  2\fb' \int \ovGplus \delta_2 R + 2\fb' \int \delta_1 \Gplus \delta_1 R 
\Biggr\},
\label{DW-tensor-2nd-order} 
\\
\delta_2 \Delta^{B} G_{\mu\nu}  &\!\!\!\!\!=\!\!\!\!\!& \left( \frac{1}{2} \delta^{\{\alpha}_\mu \delta^{\beta\}}_\nu - \frac{1}{2} \eta_{\mu\nu} \eta^{\alpha\beta} \right) \partial_\alpha \left(\int \ovGplus \delta_1 R \right)
\partial_\beta \left( \fb' \int \ovGplus \delta_1 R \right).
\label{DW-tensor-2nd-order_2} 
\eea
}
The integral
\begin{align}
\int \ovGplus \delta_1 R \equiv \int \ovGplus[x-x'] \delta_1 R[x'] \dd^4 x'
\end{align}
has already been evaluated in Eq. \eqref{1st-Green-integral} to give the local term
\begin{align}
\int \ovGplus \delta_1 R = \frac{1}{2} \hb .
\end{align}
This allows us to make the crucial observation, that the only nonlocal contribution to the $\fb''$ terms in Eqs. \eqref{DW-tensor-2nd-order} and \eqref{DW-tensor-2nd-order_2} is the last term on the second line in Eq. \eqref{DW-tensor-2nd-order}:
\begin{align}
\label{GWStressTensor___}
t_{\mu\nu} = -\frac{\fb''}{8\pi\GN} \left( \eta_{\mu\nu} \partial^2 - \partial_\mu \partial_\nu \right) I_1 + \dots
\end{align}
where 
\begin{align}
\label{I1}
I_1
&=  \int \dd^4 x' \ovGplus[x-x'] \delta_1 R[x'] \int d^4 x'' \ovGplus [x' - x'']\delta_1 R[x''] \\
&=  \int \dd^4 x' \ovGplus[x-x'] \frac{1}{2} \partial^2_{x'} \hb[x'] \frac{1}{2} \hb[x'].
\end{align}
Further note that, the only two terms proportional to $\bar{f}''$ in Eqs. \eqref{DW-tensor-2nd-order} and \eqref{DW-tensor-2nd-order_2} are both invariant under the gauge transformation $h_{\mu\nu} \to h_{\mu\nu} + \partial_{\{\mu} \xi_{\nu\}}$, induced by the infinitesimal coordinate change $x^\mu \to x^\mu + \xi^\mu$, because they both contain the gauge-invariant linearized Ricci scalar $\delta_1 \mathcal{R}$.

Now, taking the trace of the equation of motion (\ref{DeserWoodardModel_deDonder}) gives
\bea
\label{hbPDE}
\partial^2\hb = -\frac{16\pi\GN}{1+\fb - 6\fpb}\tau ,
\eea
which in turn allows us to solve $\hb$ using Eq. \eqref{GreenFunction},
\begin{align}
\label{hb_integral}
\hb[x'] = -\frac{16\pi\GN}{1+\fb - 6\fpb} \int \dd^4 x'' \ovGplus[x'-x'']\tau[x''] .
\end{align}
Inserting Eqs. \eqref{hbPDE} and \eqref{hb_integral} into Eq. \eqref{I1},
\begin{align}
\label{I1_tau}
I_1
= \frac{1}{4} \left(\frac{16\pi\GN}{1+\fb - 6\fpb}\right)^2 
\int \dd^4 x' \ovGplus[x-x'] \left(\tau[x'] \int \dd^4 x'' \ovGplus[x'-x'']\tau[x'']\right) .
\end{align}
At this order in perturbation theory, we may replace $\tau \to T \equiv \eta^{\mu\nu} T_{\mu\nu}$, because the GW contribution necessarily scales at least as $\mathcal{O}(G_\text{N})$. Furthermore, if we make the simplifying assumption that the astrophysical system is non-relativistic, then $T \approx T_{00}$; i.e., energy density dominates over pressure density. Additionally, upon replacing $\tau$ with only its matter contribution, notice the integrand is strictly zero outside the matter source. This allows us to simultaneously take the far zone and non-relativistic limits of $I_1$ readily:
\begin{align}
I_1
\approx \frac{1}{16\pi r} \left(\frac{16\pi\GN}{1+\fb - 6\fpb}\right)^2 
\int \dd^3 \vec{x}' T_{00}[t-r,\vec{x}'] \int \dd^3 \vec{x}'' \frac{T_{00}[t-r-|\vec{x}'-\vec{x}''|,\vec{x}'']}{4\pi |\vec{x}'-\vec{x}''|} .
\end{align}
The $\vec{x}'$ and $\vec{x}''$ both lie within the source, so $r \gg |\vec{x}'-\vec{x}''|$ by assumption.
\begin{align}
\label{I1_T00}
I_1
\approx \frac{1}{r} \frac{16\pi\GN^2}{(1+\fb - 6\fpb)^2} 
\int \dd^3 \vec{x}' \int \dd^3 \vec{x}'' \frac{T_{00}[t-r,\vec{x}'] T_{00}[t-r,\vec{x}'']}{4\pi |\vec{x}'-\vec{x}''|} .
\end{align}
Because $\partial^2 (\mathcal{A}[t-r]/r) = 0$ for any amplitude $\mathcal{A}$, after plugging Eq. \eqref{I1_T00} into Eq. \eqref{GWStressTensor___}, we arrive at the main result -- the only nonlocal contribution to the GW stress energy that is proportional to $\fb''$, goes as $1/r$ in the far zone location of the observer:
\begin{align}
t_{\mu\nu} 
&= \frac{\fb''}{r} \frac{2\GN}{(1+\fb - 6\fpb)^2} \left(\delta_\mu^0 - \widehat{r}^i \delta^i_\mu \right) \left(\delta_\nu^0 - \widehat{r}^j \delta^j_\nu \right) \nonumber \\
&\qquad\times
\frac{\partial^2}{\partial t^2}  
\int \dd^3 \vec{x}' \int \dd^3 \vec{x}'' \frac{T_{00}[t-r,\vec{x}'] T_{00}[t-r,\vec{x}'']}{4\pi |\vec{x}'-\vec{x}''|}  + \dots
\label{DivergentGWFlux}
\end{align}
We reiterate there is no need to compute other nonlocal terms in Eqs. \eqref{DW-tensor-2nd-order} and \eqref{DW-tensor-2nd-order_2}, because there are no other nonlocal $\fb''$ terms that could potentially cancel this $1/r$ divergent flux.

Before closing this section, we note that the other nonlocal contributions to the GW stress energy are the ones proportional to $\fb'$ on the second line in Eq. \eqref{DW-tensor-2nd-order}:
\begin{align}
\label{GWStressTensor_I2-I3}
-\frac{2\fb'}{8\pi\GN} \left( \eta_{\mu\nu} \partial^2 - \partial_\mu \partial_\nu \right) (-I_1 +I_2 + I_3) , 
\end{align}
where 
\begin{align}
\label{I2}
I_2&\equiv  \int d^4 x' \ovGplus[x-x']\delta_2 R[x'] ,
\\
\label{I3}
I_3 &\equiv  \int d^4 x' \delta_1 G[x,x'] \delta_1 R[x'] .
\end{align}
Our preliminary analysis finds that $I_2$ receives a contribution from the region of spacetime near the non-relativistic source of GWs, such that a divergent $1/r$ flux is also generated:
\be
I_2 \approx  \frac{16\pi\GN^2}{r}   \Biggl\{
\frac{2}{3(1+\fb)^2} +  \frac{5}{6(1+\fb-6\fpb)^2}
 \Biggr\}
 \int \dd^3\vec{x}' \int \dd^3 \vec{x}'' \frac{T_{00}[t-r,\vec{x}'] T_{00}[t-r,\vec{x}'']}{4\pi |\vec{x}'-\vec{x}''|} .
 \label{I2-final}
\ee 
Furthermore, $I_3$ contains extra derivatives acting on $T_{\mu\nu}$, rendering it sub-dominant with respect to $I_2$:
\begin{align}
I_3 &=-\frac{1}{1+\fb - 6\fpb} \partial_\mu \partial_{\nu} \int d^4 x' \int d^4 x'' \ovGplus[x-x''] \hb^{\mu\nu}[x'']  \ovGplus[x''-x']\tau[x'] .
\label{I_2-start}
\\
& \approx - \frac{16\pi \GN^2}{r}\frac{(1+\fb - 4\fpb)}{(1+\fb)(1+\fb - 6\fpb)^2}  \partial_0^2  \int d^3 \vec{x}' \int \dd^3 \vec{x}'' \frac{T_{00}[t-r,\vec{x}' ]T_{00}[t-r, \vec{x}'']}{4\pi} |\vec{x}' - \vec{x}''|  + \cdots
\end{align}
That is, it is unlikely that $I_3$ can cancel the divergent flux of $I_2$. Once again, we observe that the only way to avoid the $1/r$ divergent fluxes of $I_2$ and $I_3$ is to set the overall coefficient $\fb'$ in Eq. \eqref{GWStressTensor_I2-I3} to zero.

\section{Discussion}
\label{Section_Discussion}

We have examined the features of gravitational waves in the DW nonlocal gravity model \eqref{DW-action}. The linearized field equations around the flat spacetime background reveal that the gravitational wave polarizations in the Deser-Woodard model are the same as in General Relativity, except the effective Newton's constant is re-scaled. Similarly, the form of the Newtonian gravitational tidal forces exerted by a non-relativistic body is the same as that of GR, but the effective Newton's constant has a different re-scaling, i.e., $G_{\text{GW}} \neq G_{\text{tidal}}$. And, the physical asymptotically flat non-rotating black hole solution in the DW model, formed from gravitational collapse, is likely not a solution of GR. All these deviations from GR arising at linear order can be avoided by requiring  $f'[0] = 0$. 
At quadratic order the gravitational energy-momentum flux due to an isolated system turns out to scale as $1/r$, which would lead to a divergent total GW energy-momentum at infinity. This divergent flux can be avoided if we set $f''[0]=0$ in addition to $f'[0]=0$.

The GW flux result, in particular, suggests a new theoretical consistency test of modified gravity theories that are nonlocal in character. In local theories, as long as the linearized solutions go as $1/r$, the first nonlinear (i.e., quadratic) corrections to their equations-of-motion can be expected to go as $1/r^2$. Since stress-energy tensors begin at quadratic order, this would provide a finite total energy-momentum at $r = \infty$. However, the nonlocal nature of the interactions within the Deser-Woodard model give rise, at the first nonlinear oder, to terms in its equations-of-motion, Eqs. \eqref{DW-tensor-2nd-order} and \eqref{DW-tensor-2nd-order_2}, that are more akin to -- though not precisely the same as -- those encountered in the GR solution of the gravitational perturbation $h_{\mu\nu}$ sourced by the stress tensor of the GWs themselves. These latter terms are typically dubbed the first post-Minkowskian corrections to linearized GR, encountered when solving Einstein's equations at the first nonlinear order. Despite their nonlinear nature, they do produce additional $1/r$ terms. Because the Deser-Woodard model GW stress tensor contains terms of similar structure, it is therefore not surprising it too receives a $1/r$ contribution; for e.g., in Eq. \eqref{DivergentGWFlux}. 

We plan to investigate whether other nonlocal models suffer from the same issue of divergent gravitational fluxes. First, it should be noted that the $m^2 \frac{1}{\Box} R$ model of \cite{VAAS-2017} is likely unable to avoid the divergent flux because the nonlocal factor $\frac{1}{\Box} R$ generates $I_2$ in Eq. \eqref{I2-final}, which contains the $1/r$ flux. 
We are interested in looking into the $m^2 R \frac{1}{\Box^2} R$ model proposed by Maggiore and Mancarella \cite{MM-2014}, which has attracted substantial attention \cite{Maggiore-1403, Koivisto-1406, BLHBP-1408, DM-1408, Maggiore-1411, Maggiore-1602, Maggiore-1603, NAAKR-1606, Maggiore-review, Dirian-1704, BDFM-1712,Kumar-1808,Kumar-1810}.
It would be also interesting to see whether nonlocal models replacing dark matter, for instance the nonlocal MOND (MOdified Newtonian Dynamics) model \cite{DEW-1106, DEW-1405,KRSTWX-1608, TW-1804} involving a more complicated nonlocal scalar, induce divergent gravitational fluxes.

As we have already pointed out, the DW model can avoid the divergent fluxes by setting $f'[0]=0=f''[0]$. Indeed, Deser and Woodard have suggested to set $f[X]=0$ for all $X \geq 0$ in order to eliminate any deviations from GR inside gravitationally bound systems \cite{W-review, DW-2013}. The justification for this choice is as follows: 
Nonlocal modifications are thought to represent quantum corrections from infrared gravitons created during primordial inflation. Since those gravitons were of horizon scale, they have the biggest effect on large scales and no effect on small scales. In other words, the nonlocal quantum effects give large modifications on large scales and no modification on small scales, which perfectly complies with the original motivation of modifying gravity on large scales \cite{W-review, DW-2013}.  

An infinite flux of gravitational energy-momentum is such a drastic result that it prompts us to seek guidance from fundamental theory in constructing nonlocal models of modified gravitation. Presumably, a first-principles derivation of the nonlocal effective field equations through quantum loop effects during primordial inflation would lead to well-defined GW energy-momentum fluxes at infinity.
Although such a derivation, which might involve a nonperturbative resummation of loop corrections, is not currently available, there is a hint towards it from perturbative calculations. 
First, for the case of the de Sitter background, where the scale factor $a[t] = e^{Ht}$ with $H$ constant, 
the nonlocal scalar becomes  \cite{Romania},
\be
\frac{1}{\Box} R \Big|_{\text{dS}} = - 4\ln[a] + \cdots .
\ee
Second, 
one loop contributions to the graviton self-energy from a massless, minimally coupled scalar field in a locally de Sitter background induce corrections to the Newtonian potential associated with a static point mass resulting in a secular decrease of the Newton's constant, proportional to $-\ln[a]$ at late times \cite{PPW-1510},
\be
   \GN \rightarrow \GN \bigg(1-\frac{1}{30\pi}\frac{ \hbar \GN H^2}{c^5}\ln[a] + \cdots \bigg). 
\label{secular-decrease}   
\ee
Also, note that primordial inflation is thought to be very close to the de Sitter background. Therefore, this $\ln[a]$ correction seems to indicate a connection between nonlocal gravity and quantum infrared loop corrections during primordial inflation.

\section*{Acknowledgements}
\label{Section_Acknowledgements}

We are grateful for correspondence on this subject with Richard Woodard. 
YZC wishes to thank the hospitality offered by CEICO, when he visited SP over summer 2018 to initiate the current project.
YZC is supported by the Ministry of Science and Technology of the R.O.C. under the grant 106-2112-M-008-024-MY3.
SP is grateful for the hospitality provided by NCU, where the main computation was performed.    
SP acknowledges the Taiwan Travel Fellowship received from the Czech Academy of Sciences (CAS), which supported her to visit institutions in Taiwan (including NCU) for three weeks in August 2018. 
The work of SP was supported by the European Research Council 
under the European Union's Seventh Framework Programme 
(FP7/2007-2013)/ERC Grant No. 617656, ``Theories and Models of the 
Dark Sector: Dark Matter, Dark Energy and Gravity".

\appendix

\section{Perturbative Expansions of Nonlocal Terms}
\label{Section_ExpansionOfNonlocalTerms}

In this section, the nonlocal factors in Eq. \eqref{DW-EoM} are expanded as up to second order in the power series of $h_{\mu\nu}$:
{\allowdisplaybreaks
	\bea
	X &\!\!\!\equiv\!\!\!&  \frac{1}{\Box} R = \delta_1 X + \delta_2 X =  \int \ovGplus \delta_1 R +  \int \left( \delta_1 \Gplus \delta_1 R + \ovGplus \frac{1}{2} h \delta_1 R + \ovGplus \delta_2 R  \right),
	\\
	f[X] &\!\!\!\equiv\!\!\!& f\left[ \frac{1}{\Box} R \right]  =\fb + \delta_1 f + \delta_2 f = \fb + \fb' \delta_1 X + \fb' \delta_2 X + \frac{1}{2}\fb'' (\delta_1 X)^2 
	\\
	&\!\!\!=\!\!\!& \fb + \fb' \int \ovGplus \delta_1 R  + \fb' \int \left( \delta_1 \Gplus \delta_1 R + \ovGplus \frac{1}{2} h \delta_1 R + \ovGplus \delta_2 R  \right) + \frac{1}{2}\fb'' \Bigl(\int  \ovGplus \delta_1 R \Bigr)^2, 
	\\
	f'[X] &\!\!\!\equiv\!\!\!& f'\left[ \frac{1}{\Box} R \right]  =\fb' + \delta_1 f' + \delta_2 f' = \fb' + \fb'' \delta_1 X + \fb'' \delta_2 X + \frac{1}{2}\fb''' (\delta_1 X)^2 
	\\
	&\!\!\!=\!\!\!& \fb' + \fb'' \int \ovGplus \delta_1 R  + \fb'' \int \left( \delta_1 \Gplus \delta_1 R + \ovGplus \frac{1}{2} h \delta_1 R + \ovGplus \delta_2 R  \right) + \frac{1}{2}\fb''' \Bigl(\int  \ovGplus \delta_1 R \Bigr)^2, 
	\\
	U &\!\!\!\equiv\!\!\!& \frac{1}{\Box} \left( R f'\left[\frac{1}{\Box} R\right] \right) = \frac{1}{\Box} \left( R f'[X] \right) = \delta_1 U + \delta_2 U
	\\
	&\!\!\!=\!\!\!& \int \ovGplus \delta_1 R  \fb'
	+  \int \Bigl(\delta_1 \Gplus \delta_1 R  \fb' + \ovGplus \frac{1}{2} h \delta_1 R+  \ovGplus \delta_2 R \fb' + \ovGplus \delta_1 R \delta_1 f'  \Bigr) 
	\\
	&\!\!\!=\!\!\!&  \fb' \int \ovGplus \delta_1 R 
	+   \fb' \int \Bigl(\delta_1 \Gplus \delta_1 R + \ovGplus \frac{1}{2} h \delta_1 R +  \ovGplus \delta_2 R   \Bigr) 
	+  \fb'' \int \ovGplus \delta_1 R \int  \ovGplus \delta_1 R.
	\eea
}
Here $\fb = f[0],  \fb' = f'[0], \fb'' = f''[0]$ since the argument $\osq^{-1} \ovr = 0$ and the retarded Green's function is expanded as
\be
\Gplus  = \ovGplus + \delta_1 \Gplus + \mathcal{O}[h^2].
\ee
Then, the modified Einstein tensor at first order in perturbations on the Minkowski background is 
\bea
\label{DW-tensor-1st-order}
\delta_1 \Delta^{A} G_{\mu\nu}  &\!\!\!\!\!=\!\!\!\!\!& 
\delta_1 G_{\mu\nu} \fb + 2(\eta_{\mu\nu} \partial^2 - \partial_\mu \partial_\nu)\fb' \int \ovGplus \delta_1 R, \\
\delta_1 \Delta^{B} G_{\mu\nu}  &\!\!\!\!\!=\!\!\!\!\!& 0 .
\eea
The modified Einstein tensor at second order is 
\bea
\delta_2 \Delta^{A} G_{\mu\nu}  
&\!\!\!\!\!=\!\!\!\!\!& \delta_2 G_{\mu\nu} \fb+ 2\delta_1 (G_{\mu\nu} +g_{\mu\nu} \Box - \nabla_\mu \nabla_\nu ) \fb' \int \ovGplus \delta_1 R 
+ ( \eta_{\mu\nu} \partial^2 - \partial_\mu \partial_\nu) \Biggl\{
\frac{1}{2}\fb'' \left(\int  \ovGplus \delta_1 R \right)^2
\nn \\ 
&& + \fb'' \int \ovGplus \delta_1 R \int  \ovGplus \delta_1 R  + \fb' \int \ovGplus h \delta_1 R +  2\fb' \int \ovGplus \delta_2 R + 2\fb' \int \delta_1 \Gplus \delta_1 R 
\Biggr\},
\\
\delta_2 \Delta^{B} G_{\mu\nu}  &\!\!\!\!\!=\!\!\!\!\!& \left( \frac{1}{2} \delta^{\{\alpha}_\mu \delta^{\beta\}}_\nu - \frac{1}{2} \eta_{\mu\nu} \eta^{\alpha\beta} \right) \partial_\alpha \left(\int \ovGplus \delta_1 R \right)
\partial_\beta \left( \fb' \int \ovGplus \delta_1 R \right).
\eea

\section{Gravitational Green's Function: de Donder Gauge}
\label{Section_GreenFunction}

In this section we will obtain the Green's function to the wave operator on the left hand side of Eq. \eqref{DeserWoodardModel_deDonder}.

{\bf Fourier spacetime solutions} \qquad We may write Eq. \eqref{DeserWoodardModel_deDonder} in Fourier spacetime.
\begin{align}
\label{DeserWoodardModel_deDonderFourier}
K_{\mu\nu}^{\phantom{\mu\nu}\alpha\beta} \widetilde{\hb}_{\alpha\beta}[k] &= 8\pi\GN \widetilde{\tau}_{\mu\nu}[k] , \\
K_{\mu\nu}^{\phantom{\mu\nu}\alpha\beta} &\equiv \frac{1}{2} \left( 1 + \fb \right) \frac{1}{2} \delta_\mu^{\{\alpha} \delta_\nu^{\beta\}} k^2
- \fpb \left( \eta_{\mu\nu} k^2 - k_\mu k_\nu \right) \eta^{\alpha\beta}, \\
k^2 &\equiv k_\sigma k^\sigma .
\end{align}
We see that, if a $K^{-1}$ can be found that satisfies
\begin{align}
K_{\mu\nu}^{\phantom{\mu\nu}\rho\sigma} (K^{-1})_{\rho\sigma}^{\phantom{\rho\sigma}\alpha\beta} 
= \frac{1}{2} \delta_\mu^{\{\alpha} \delta_\nu^{\beta\}} ,
\end{align}
then
\begin{align}
\widetilde{\hb}_{\mu\nu}[k] 
= 8\pi\GN (K^{-1})_{\mu\nu}^{\phantom{\mu\nu}\alpha\beta}[k] \widetilde{\tau}_{\alpha\beta}[k] .
\end{align}
In fact, a direct calculation would yield
\begin{align}
-(K^{-1})_{\mu\nu}^{\phantom{\mu\nu}\alpha\beta}
&= \frac{2}{1+\fb} \frac{1}{2} \delta_\mu^{\{\alpha} \delta_\nu^{\beta\}} \frac{1}{-k^2}
+ \frac{4 \fpb}{(1+\fb)(1+\fb-6\fpb)} \frac{\eta^{\alpha\beta}}{-k^2} \left( \eta_{\mu\nu} - \frac{k_\mu k_\nu}{k^2} \right) .
\end{align}
The reason for the $-$ sign is to highlight that $1/(-k^2) = 1/\partial^2$. 

Now, the retarded Green's function for the wave operator in 4D Minkowski is
\begin{align}
\frac{1}{\partial^2}[z \equiv (t,\vec{z})] \equiv \overline{G}^+[z] 
&= \int_{-\infty+i0^+}^{+\infty+i0^+} \frac{\dd k_0}{2\pi} \int_{\mathbb{R}^3} \frac{\dd^3 \vec{k}}{(2\pi)^3} e^{i\vec{k}\cdot\vec{z}} \frac{e^{-i k_0 t}}{-k_0^2+\vec{k}^2} \\
&= \frac{\delta[t-r]}{4\pi r}, \qquad\qquad r \equiv |\vec{z}| \\
&= \frac{\Theta[t]}{4\pi} \delta[\sbar], \qquad \sbar \equiv \frac{t^2-r^2}{2} .
\end{align}
We also have the object
\begin{align}
g_{\mu\nu}[z] 
&\equiv \int_{-\infty+i0^+}^{+\infty+i0^+} \frac{\dd k_0}{2\pi}
\int_{\mathbb{R}^3} \frac{\dd^3 \vec{k}}{(2\pi)^3} e^{i\vec{k}\cdot\vec{z}} \frac{e^{-ik_0 t}}{-k^2} \frac{k_\mu k_\nu}{k^2},
\qquad k^2 \equiv k_0^2-\vec{k}^2 \\
&= \frac{1}{2} i \frac{\partial}{\partial z^\mu} \int \frac{\dd^4 k}{(2\pi)^4} e^{-ik \cdot z} \frac{\partial}{\partial k^\nu} \frac{1}{k^2} .
\end{align}
Assuming it is alright to integrate-by-parts the $\partial/\partial k^\nu$,
\begin{align}
g_{\mu\nu}[z] 
&= -\frac{1}{2} i \frac{\partial}{\partial z^\mu} \int \frac{\dd^4 k}{(2\pi)^4} (-iz_\nu) \frac{e^{-ik \cdot z}}{k^2} \\
&= \frac{1}{2} \frac{\partial}{\partial z^\mu} \left\{ z_\nu \int \frac{\dd^4 k}{(2\pi)^4} \frac{e^{-ik \cdot z}}{-k^2} \right\} \\
&= \frac{1}{2} \frac{\partial}{\partial z^\mu} \left\{ z_\nu \overline{G}^+[z] \right\} \\
&= \frac{\Theta[t]}{8\pi} \left( \eta_{\mu\nu} \delta[\sbar] + z_\mu z_\nu \delta'[\sbar] \right) .
\end{align}
As a check of this result, we note from its Fourier representation that $\eta^{\mu\nu} g_{\mu\nu} = \overline{G}^+$. From the final position spacetime result,
\begin{align}
\eta^{\mu\nu} g_{\mu\nu}
&= \frac{\Theta[t]}{8\pi} \left( 4 \delta[\sbar] + 2 \sbar \delta'[\sbar] \right) .
\end{align}
Using the identity $\sbar \delta'[\sbar] = -\delta[\sbar]$ we indeed obtain $\Theta[t] (4\pi)^{-1} \delta[\sbar] = \overline{G}^+$. It is also possible to arrive at this result by first tackling the Fourier integral
\begin{align}
g_2[z] \equiv \int_{\text{ret}} \frac{\dd^4 k}{(2\pi)^4} \frac{e^{-ik \cdot z}}{-(k^2)^2} .
\end{align}
The desired result is then $g_{\mu\nu} = -\partial_\mu \partial_\nu g_2[z]$.

The linearized Deser-Woodard graviton propagator contains the Fourier integral
\begin{align}
\text{DW}_{\mu\nu} = \int \frac{\dd^4 k}{(2\pi)^4} \frac{e^{-ik\cdot z}}{-k^2} \left( \eta_{\mu\nu} - \frac{k_\mu k_\nu}{k^2} \right) .
\end{align}
We have
\begin{align}
\text{DW}_{\mu\nu}[z] 
&= \frac{\Theta[t]}{4\pi} \left( \eta_{\mu\nu} \delta[\sbar] - \frac{1}{2} \frac{\partial}{\partial z^\mu} \left\{ z_\nu \delta[\sbar] \right\} \right) \\
&= \frac{\Theta[t]}{8\pi} \left( \eta_{\mu\nu} \delta[\sbar] - z_\mu z_\nu \delta'[\sbar] \right) .
\end{align}

\end{document}